\documentclass{article}
\usepackage{amsmath,amsfonts,amssymb,latexsym} 
\usepackage{graphicx}
\usepackage[]{epsfig}
\usepackage{bm}
\usepackage{stmaryrd}
\newcommand{\beq}{\begin{equation}}
\newcommand{\eeq}{\end{equation}}
\newcommand{\beqd}{\begin{displaymath}}
\newcommand{\eeqd}{\end{displaymath}}
\newcommand{\beqa}{\begin{eqnarray}}
\newcommand{\eeqa}{\end{eqnarray}}

\newcommand{\comment}[1]{}

\newcommand{\Tr}{{\rm Tr}\,}

\begin{document}

\title{Quasi-equilibrium in glassy dynamics: an algebraic view}

\author{Silvio Franz\\ {\small Laboratoire de Physique Th\'eorique et Mod\`eles
    Statistiques,} \\ {\small CNRS et Universit\'e Paris-Sud 11,
    UMR8626, B\^at. 100, 91405 Orsay Cedex, France}\\
\\
Giorgio Parisi\\
{\small Dipartimento di Fisica, INFN -- Sezione di Roma I, IPFC-CNR --
  UOS Roma}\\ {\small Sapienza Universit\`a di Roma, P.le Aldo Moro 2,
  I-00185 Roma, Italy}}


\maketitle
\begin{abstract}
  We study a chain of identical glassy systems in a constrained
  equilibrium where each bond of the chain is forced to remain at a
  preassigned distance to the previous one.  We apply this description
  to Mean Field Glassy systems in the limit of long chain where each
  bond is close to the previous one. We show that in specific
  conditions this pseudo-dynamic process can formally describe real
  relaxational dynamics for long times. In particular, in mean field spin glass 
  models we can recover in this way the equations of Langevin
  dynamics in the long time limit at the dynamical
  transition temperature  and below.  We interpret the formal identity as an
  evidence that in these situations the configuration space is
  explored in a quasi-equilibrium fashion.  Our general formalism, that
  relates dynamics to equilibrium, puts slow dynamics in a new
  perspective and opens the way to the computation of new dynamical
  quantities in glassy systems.
\end{abstract}

\section{Introduction}
Glassy dynamics is often described as a quasi equilibrium process.
Phase space exploration is depicted as walk from a metastable state to
another, the choice of which for large times is dictated by
generalized equilibrium conditions.  This picture has been used in the
past to interpret slow dynamics for liquids and glasses in equilibrium
and aging conditions \cite{FrVir,Kurchetal}, and more recently for
studying dynamical fluctuations in equilibrium terms using constrained
Boltzmann-Gibbs measures \cite{FPRR}.  Theoretical evidence in support
of this view comes from the emergence of effective temperatures in
glassy dynamics \cite{CuKuPe}, the coincidence of certain static and
dynamic quantities in the long time limit\cite{FMPP} and a detailed
analysis of the response properties during aging dynamics
\cite{FrVir}. In addition, numerical simulations of simple models
\cite{Kurchetal} and realistic systems \cite{Sciortino} agree with
this picture.

Despite the appeal of this picture and the many arguments that can be
bring to its support, a formal description of glassy dynamics in
equilibrium terms is missing, with the consequence that many 
quantities of dynamical interest as for example the entropy of 
the trajectories have not been computed even in the simplest mean-field 
models. 

In this note we would like to fill this gap by establishing a general
methodology allowing to test the dynamical quasi-equilibrium
hypothesis at least at the mean field level. To this aim we employ the
method of constrained equilibrium measure \cite{FP} and consider a
chain of replicas of the system under study, each one in constrained
equilibrium with respect to the previous one.  Such construction was used
for chains of length 2 in \cite{FP} and of length 3 in \cite{irene},
as a tool to probe the configuration space of glassy systems. The
generalization to an arbitrary number $L$ of bonds in the chain was
sketched in \cite{JFK} but no computations based on it were
presented. Recently it has been suggested that large values of $L$
might be necessary to adiabatically follow metastable states in
temperature \cite{FloLenka} and clarify some anomalies encountered in
the $L=2,3$ case. Progress has been made by Krzakala and Zdeborova in
treating this problem with the cavity method for finite $L$
\cite{FloLenka_private}. We will see that the interest of this
construction goes much beyond the problem of following states.  Here
we employ the replica method and concentrate on fully connected spin
glass models, and show how the $L\to \infty$ case relates to slow
glassy dynamics.

\section{A Markov Chain of replicas}
Given a physical macroscopic system with configurations labeled by $S$, $S'$
etc. subject to some Hamiltonian $H(S)$, 
and given a notion of similarity $q(S,S')$ between
configurations, we consider a linear chain of $t$ copies
such that:

 1) the first configuration is chosen with Boltzmann probability
at temperature $T_1$
\begin{eqnarray}
P(S_1)=\frac 1 Z \exp\left(-\beta_1 H(S_1)\right),
\end{eqnarray}
2) for any integer $s>1$ the $s+1$-th copy is drawn from the
Boltzmann-Gibbs measure at temperature $T_{s+1}$ (that may or not 
depend on $s+1$) with a chain constraint that $q(S_s,S_{s+1})$ 
is fixed to some preassigned values ${\widetilde C}_c(s+1,s)$, namely 
\begin{eqnarray}
M(S_{s+1}|S_{s})=\frac 1 {Z_{s+1}(S_s)} \exp\left(-\beta_{s+1} 
H(S_{s+1})\right )\delta(q(S_s,S_{s+1})-{\widetilde C}_c(s+1,s)).
\end{eqnarray}
Notice that the conditional probability kernel $M$ for fixed ${\widetilde
  C}_c(s+1,s)$ mathematically defines a Markov chain, where the 
probability of a trajectory is given by
\begin{eqnarray}
P(S_t,S_{t-1},...S_1)=\prod_{s=1,L-1}M(S_{s+1}|S_{s})P(S_1).
\label{chain}
\end{eqnarray}
Such a general chain construction was mentioned in \cite{JFK}, but to
our knowledge it has not been employed in actual calculations or
theories. We will often refer to the chain (\ref{chain}) as Boltzmann
pseudo-dynamics and we call the variables $s$, etc., times in the
following.  In order to understand the physical property of the chain,
we would like to study the free-energy of the ``last replica'',
\begin{eqnarray}
F(L)=-T_{L}\sum_{S_1,...,S_{L-1}}P(S_{L-1},S_{L-2},...S_1)\log Z_{L}(S_{L-1}), 
\label{FE}
\end{eqnarray}
as a function of the number of bonds in the chain $L$. Although this
free-energy will be the starting point of our analysis, we 
concentrate in this paper on the properties of the measure
(\ref{chain}), through mean-field analysis based on the replica method. 

As we stated in the introduction, 
a full analysis has been achieved in the cases
$L=2$ \cite{FP} and $L=3$ \cite{irene} with the purpose of
investigating the structure of metastable states and the barrier
separating them (in these case the temperatures are equal along the
chain) and the properties of the metastable states when cooled and the
temperatures depend on $s$. It has been recently remarked that in
order to explain certain anomalies found for $L=2$ in the case of
different temperatures, the general construction with arbitrary $L$
might be needed \cite{FloLenka}.  Unfortunately, the complexity of the
computation strongly increases with total number of steps $L$ involved
in the Markov chain. In this note we show that simple expressions can
be obtained in the limit where the total number of steps in the chain
goes to infinity: differences along neighboring bonds in the chain become
small and time becomes continuous.

\section{The replica algebra}

The problem  of analyzing the 
chain (\ref{chain}) can be addressed using the replica
method starting from the computation of the free-energy (\ref{FE}). 
One needs to replicate each of the configuration $S_s$ a
number of time $n(s)$ in principle different for each $s$ and consider
at the end the limit $n(s)\to 0$; the free-energy (\ref{FE}) being
 associated to the term of order $n(L)$ of the resulting expression.  
We will call the replicas $S_s^a$ 
with the convention that replica indexes associated to the index $s$
run from 1 to $n(s)$.
Denoting by $Q_{a,b}(s,u)=q(S_s^a,S_u^b)$ the overlap between two 
replicas $S_s^a$ and $S_u^b$, the chain constraint reads 
\begin{eqnarray}
Q_{a,1}(s+1,s)={\widetilde C}_c(s+1,s),
\label{constraint}
\end{eqnarray}
in words, all replicas $a=1,...,n(s+1)$ at time $s+1$ are constrained
to have a fixed overlap ${\widetilde C}_c(s+1,s)$ with replica number
1 at time $s$. The constraint (\ref{constraint}) can be imposed
through Lagrange multipliers $\nu(s)$. In this first paper we will
only deal with cases in which the constraints ``do not make work'' and
$\nu(t)=0$, for which we will show that remarkable solutions exist.
Future work it is planned to deal with the fully constrained case.

{\bf Mean Field Spin Glasses} In mean field spin glasses the
free-energy is obtained as a saddle point over the time-replica matrix
order parameter $Q_{a,b}(s,u)=\langle q(S_s^a,S_u^b) \rangle$. As usual, 
we need an ansatz for the replica matrix that allows the analytic continuation
to $n(s)\to 0$. In this paper we are interested to
the time structure of the matrix and we stick to a replica symmetric
 ansatz as far as the 
``$a,b$'' indexes are concerned. The
generalization to replica symmetry breaking (RSB) is straightforward and
does not pose any particular problem of
principle. In fact  we expect  RSB to be crucial in many applications and it 
will be studied elsewhere.

The form of the constraint (\ref{constraint}), symmetry considerations
and the experience gained in \cite{FP} and \cite{irene} suggests to
consider matrices $Q_{a,b}(s,u)$ that for $s\ne u$ depend only on the
index $a$ if $s<u$ and depend only on $b$ if $u<s$.  In the case of
$s=u$ the matrix $Q_{ab}(t,t)$ is assumed to have the usual
structure. Such a general scheme can easily incorporate Replica
Symmetry Breaking and this has been considered in the special cases of
$L=2,3$ in \cite{FP} and \cite{irene}.  In this paper we will limit
ourselves to the replica symmetric case where the most general matrix
can take the form
\begin{eqnarray}
Q_{a,b}(s,u)=&& C(s,u)+
[\widetilde{C}(s,u)-C(s,u)]\Theta_>(s-u)\delta_{b,1}+ \label{RS}\\\nonumber
&&[\widetilde{C}(u,s)-C(u,s)]\Theta_>(u-s)\delta_{a,1} +
[\widetilde{C}(u,u)-C(u,u)]\delta_{s,u}\delta_{a,b},
\end{eqnarray}
where $C(s,u)$ and $\widetilde{C}(s,u)$ are symmetric functions of
their arguments. Having defined $\Theta_>(s)=1$ if $s>0$ and zero
otherwise.  In Ising of spherical models $\widetilde{C}(u,u)=1$. We
prefer not to specify the value of $\widetilde{C}(u,u)=1$ at this
level to simplify the analysis of functions of the replica matrix.  We
remark that the Markov chain structure of (\ref{chain}) implies that in the $n(s)\to 0$
limit the saddle point equations must have a causal structure, and the
equations for $C(s,u)$ and ${\widetilde C}(s,u)$ should not contain
values of the functions at later times. One can see that this
causality property is respected whenever $Q_{a,b}(s,u)$ only depends
on index $a$ if $s<u$ and on the index $b$ if $u<s$.

In this way, for example, the equations for $C(1,1)$ is the usual RS
equation for the Edwards-Anderson parameter in a single system. 
We will here consider the so called
``annealed cases'' where the solution is $C(1,1)=0$ but the formalism in not
specific to this case. It is reasonable (and consistent with the equations) 
in the annealed case to take
$C(1,s)=0$, choice that we will adopt in the following. In this paper
we concentrate on the structure of the saddle-point equations and some
physical implications of their solutions, letting the study of the
free-energy to future work.

In many models (e.g in the spherical spin glass model or in the Ising
Sherrington-Kirkpatrick model near to the phase
transition\footnote{This is true up to the order $(T_{c}-T) ^{10}$ 
in an expansion in $T_c-T$!})
the only non trivial operation that is needed in order to solve the
saddle point equations is the product of two replica matrices. This is
well studied in the case of the standard hierarchical structure, where
the matrix is parametrized in terms of a function $q(x)$ \cite{PARISI}.

We are now interested to find out a simple expression for this
product. As we have already mentioned we want to study the limit where
the discrete time $t$ goes to infinity and the Markov chain collapses
onto a Markov process. To this end it is convenient to evaluate the
product between two matrices $Q_A$ and $Q_B$ which have the structure
(\ref{RS}), i.e. they are parametrized in terms of functions
$C_{g}(s,u)$ and ${\widetilde C}_{g}(s,u)$ with $g=A,B$ respectively.

It turns out that a good long chain limit $L\to\infty$ is obtained if
one suppose that $C(s,u)$ tends to continuous function of $s/L$ and
$u/L$ and for $s\ne u$ and for $s,u\ne 1$, one has
$\widetilde{C}(s,u)-C(s,u)=O(1/L)$. We therefore abandon the discrete
time: without causing confusion we change notation and from now on
the variables $u,s$, etc. will denote continuous variables taking
values in some interval $[0,t_{max}]$, (without necessarily normalizing the
final point to $t_{max}$ to 1).  We then define the function
\begin{eqnarray}
T_u R(s,u)\; du=\Theta(s-u)(\widetilde{C}(s,u)-C(s,u)).
\label{risp}
\end{eqnarray}
that we call response function,
we will see later that this name is non-abusive.

Denoting therefore by $Q_C$ the product between $Q_A$ and $Q_B$, one
finds that $Q_C$ is consistently parametrized by functions $C_C(s,u)$
and $R_C(s,u)$. In the continuous limit a careful computation\footnote{In fact this computation can be fully automatized implementing the matrix multiplication on an algebraic manipulation software.} shows
that  in the limit in which all the $n(s)$ go to zero, 
these functions verify the relations:
\begin{eqnarray}
C_C(t,r)= & &
\int_0^r ds\; 
C_A(t,s) T_s R_B(r,s)+ \int_0^t ds\; T_s R_A(t,s)C_B(r,s)\\ \nonumber
& &+[\widetilde{C}_A(t,t)-C_A(t,t)]C_B(t,r)+C_A(t,r)[\widetilde{C}_B(r,r)-C_B(r,r)] \\\nonumber
&& +C_A(t,0)C_B(r,0),\\
R_C(t,r)=& &
\int_r^t ds\; R_A(t,s) R_B(s,r)\\ \nonumber
 & &+\beta_r R_A(t,r)[\widetilde{C}_B(r,r)-C_B({r,r})]
+\beta_r [\widetilde{C}_A(t,t)-C_A(t,t)]R_B(t,r),\\ \nonumber
\widetilde{C}_C(t,t)&=&
\int_0^t ds\; 
C_A(t,s) T_s R_B(t,s)+ T_s R_A(t,s)C_B(t,s)\nonumber\\
&&+\widetilde{C}_A(t,t)\widetilde{C}_B(t,t)-{C}_A(t,t){C}_B(t,t)
\\ \nonumber
&& +C_A(t,0)C_B(t,0).\\
 \label{replicasproduct}
\end{eqnarray}
The alerted reader will recognize the similarity of eq.ns
(\ref{replicasproduct}) with the convolution of two functions in the
supersymmetric formalism used in the Langevin relaxational dynamics
starting from random initial conditions \cite{MEZARD}.  In that 
context one defines
the supersymmetric correlation function
\begin{eqnarray}
Q(t, \theta;s,\theta')= C(t,s)+\theta T_t R(s,t)+\theta' T_s R(t,s),
 \label{Qsusy}
\end{eqnarray}
where $\theta$ and $\theta'$ are (commuting) Grassmanian variables. The convolution 
between two functions of that type
\begin{eqnarray}
Q_C(t, \theta;s,\theta')=\int du\; d\theta''\; Q_A(t, \theta;u,\theta'')Q_B(u, \theta'';s,\theta')
\end{eqnarray}
is still a function of the same form with 
\begin{eqnarray}
C_C(t,r)= & &
\int_0^r ds\; 
C_A(t,s) T_s R_B(r,s)+ \int_0^t ds\; T_s R_A(t,s)C_B(r,s)\\ \nonumber
R_C(t,r)=& &
\int_r^t ds R_A(t,s) R_B(s,r).\\ \nonumber
 \label{susyproduct}
\end{eqnarray}
We see that our replica product equals the susy one if $C(t,0)=0$ and
${\widetilde C}(t,t)=C(t,t)$. In this respect we notice that the term
containing $C(t,0)$ comes as consequence of choosing the first replica
in equilibrium, and would disappear if for example $T_1\to\infty$, or 
if the memory of the initial condition would be lost. 
As the matter of fact, if one considers Langevin dynamical relaxation
starting from an equilibrium initial condition the supersymmetric
product is modified and that very additional term appears \cite{BFP}.
The other additional term comes from a fundamental difference between
real dynamics and Boltzmann pseudo-dynamics. In real dynamics, short
time scales are dominated by fast relaxation processes and one cannot
say in any sense that the vicinity of a given instantaneous
configurations is explored according to the Boltzmann weight. This property 
can only hold on large time scales. By
contrast, pseudo-dynamics samples according to Boltzmann by
construction.  Notice that the minimal distance between subsequent
bonds in the chain is ${\widetilde C}(t,t)-C(t,t)$, which one can expect to be
macroscopic.  In situations where time scale separation occurs, and
the fast part of the dynamics is seen as instantaneous by the slow one,
the additional terms that we find in the replica product 
exactly coincide with the ones that couple the
slow part to the dynamics to the fast one in Langevin dynamics. 
We will see that in fact
this property implies that the resulting equations have the celebrated
property of invariance \cite{russi} under time reparametrizations
$t\to h(t)$ for monotonous functions $h(t)$.
 

It is remarkable that the replicas algebra in the limit of continuous
time and vanishing $n(s)$ is isomorphic to the supersymmetric algebra
of the dynamics. In fact, this longly sought isomorphism
\cite{leticia} roughly reduces to the correspondence
\begin{eqnarray}  
\delta_{b,1} \;ds \to \theta 
\end{eqnarray}  
and to  neglecting systematically terms of order $ds^2$. 

Albeit much more
complicated, a formula for the the matrix product can be written for finite
$n(s)$. The resulting structure constitutes a deformation of the susy
algebra, whose formal properties would be interesting to study.

Having established the product algebras for the replicas, 
we segue into the computation of the equations for the spherical model.

\section{ Spherical $p$-spin models} 

The replica analysis, as well as the study of dynamics are simplified
in mean field spherical models \cite{crisantisommers}. In these models
the spins $S_i$ verify the spherical constraint $\sum_{i=1}^N S_i^2=N$
and the Hamiltonian $H(S)$ is a random Gaussian function of the
configurations with covariance
\begin{eqnarray}
\label{corre}
\overline{H(S)H(S')}=Nf(q(S,S')),
\end{eqnarray} 
where the function $f(q)$ is in general chosen to be a polynomial with
positive coefficients. We concentrate here on functions $f$ leading to
``one step'' RSB in statics, as for example one finds in the case 
$f(q)=q^p$ with $p\geq 3$.

The replica analysis of the model shows that the replicated 
free-energy as a function 
of a generic replica matrix can be 
written as \cite{crisantisommers}
\begin{eqnarray}
-\beta F[Q]=\frac{1}{2}\sum_{a,b,s,u}\beta_s\beta_u 
f(Q_{a,b}(s,u))+\frac 1 2 \Tr \log Q
-\frac 1 2 \Tr \mu (Q-C_c),
\end{eqnarray}
where the last term is needed to enforce the constraints
(\ref{constraint}), as announced we consider the case of $\nu(t)=0$, 
and 
$\mu_{a,b}(t,s)=\mu(t)\delta_{t,s}\delta_{a,b}$, where $\mu(t)$ enforces the spherical constraint at all times. 
The saddle point equations read 
\begin{eqnarray}
\frac{\beta_s \beta_u}{2} f'(Q_{a,b}(s,u))+\frac 1 2 Q^{-1}_{a,b}(s,u)
-\frac 1 2  \mu(s)\delta_{a,b}\delta(s-u) =0. 
\end{eqnarray}
In order not to need to invert $Q$ we can just multiply by $Q$ and get 
\begin{eqnarray}
\frac{1}{2} \sum_{u=1}^L\sum_{b=1}^{n(u)} 
\beta_s \beta_u f'(Q_{a,b}(s,u))Q_{b,c}(u,v)+\frac 1 2 \delta_{a,b}\delta_{s,v}
-\frac 1 2   \mu(s)Q_{a,c}(s,v) =0. 
\end{eqnarray}
Inserting (\ref{replicasproduct}) we get the equations:
\begin{eqnarray}
\mu(t)C(t,u)
= &&\beta_t \int_0^u ds\; 
f'(C(t,s)) R(u,s)+ \beta_t \int_0^t ds\; f''(C(t,s)) R(t,s)C(u,s)
\nonumber\\ 
&& +\beta_t \beta_u (f'(1)-f'(C(t,t)))C(t,u)+\beta_t \beta_u
f'(C(t,u))(1-C(u,u)) \\\nonumber
&& +\beta_t \beta_u f'(C(t,0))C(u,0),\\\nonumber
\mu(t)R(t,u)
= & &
\beta_t\int_u^t f''(C(t,s))R(t,s)ds R(s,u)\\ \nonumber
  &&+\beta_t f''(C(t,u))R(t,u)(1-C({u,u}))
+\beta_t (f'(1)-f'(C(t,t)))R(t,u),\\ \nonumber
  \mu(t)=&&
 T_t +\beta_t^2(f'(1)-f'({C}(t,t)){C}(t,t))\\ 
 &&+\beta_t \int_0^t ds\; 
\left( f'(C(t,s)) R(t,s)+ f''(C(t,s)) R(t,s)C(t,s)\right)+\beta_t^2
f'(C(t,0))C(t,0), \nonumber\\
\nonumber
\end{eqnarray}
where we used the condition ${\widetilde C}(t,t)=1$. 
As announced these equation are for constant $\beta_t=\beta$ invariant under 
time reparametrization. Notice that we can in fact
use this invariance to chose the dependence of $\beta_t$
on the time.  These equations and some of their solutions are well
known \cite{crisantihornersommers,Gotze,CuKu,BFP}. 
In fact non-trivial solutions have been found in two cases, describing:
\begin{enumerate}
\item The equilibrium alpha relaxation process for constant
  temperature $T\to T_d$ ($T_d$ is the dynamical transition
  temperature of the model) \cite{Gotze}. In this case one chooses the
  function $C(t,s)$ and $R(t,s)$ to be time translation invariant and
  verifying the fluctuation dissipation relation $R(t-s)=\beta
  \frac{\partial C(t-s)}{\partial s}$.
\item The slow part of aging relaxation starting from a
  non-equilibrium condition, for $T<T_d$ \cite{CuKu}. This situation
  can be achieved in our formalism if we take a very high value of
  $T_1$ and later a constant temperature $T$, and supposing
 loss of memory of the initial condition
  $C(t,0)=0$. One finds then a family of solutions of the kind
  $C(t,s)=C(h(s)/h(t))$ (if $t>s$), with the response verifying 
the modified fluctuation dissipation relation 
$R(t,s)=\beta x  \frac{\partial C(t,s)}{\partial s}$ with $x\in [0,1]$. 
\end{enumerate}
In both cases the dynamics is critical. Indeed marginal stability,
physically associated to vanishing of free-energy barriers, appears as
a necessary condition for having non-trivial solutions where $C(t,s)$
actually depends on time, and in last analysis for the equivalence of
slow real dynamics and Boltzmann pseudo-dynamics.

\section{The SK model} 
The reader may wander at this point if our findings are specific of
models where only products and functions of the replica 
matrix elements are important. In this section
we will study the Sherrington-Kirkpatrick model (with Ising spins) and
show that the equivalence with long time dynamics still holds.
The Hamiltonian of the model is a Gaussian function as in the previous section 
with a correlation function specified by (\ref{corre}) with $f(q)=q^2/2$
For simplicity we will consider the case of infinite initial 
temperature $T_1\to\infty$ and constant temperature $T_s=T$ for $s>1$. 
We will not include in the analysis a low temperature
equilibrium initial condition. Contrary to the p-spin case above
$T_{stat}$ this would require replica symmetry breaking, which is not
the main emphasis here.

Quoting from \cite{MPV}, we write the replicated partition function of
the model (up to irrelevant terms) as
\begin{eqnarray}
Z_{Rep}=&&{\rm s.p.}\; e^{-N \frac{\beta^2}{2} \sum_{\alpha<\beta} Q^2_{\alpha,\beta}}\times \zeta[Q]^N \nonumber \\
&& \zeta[Q]=\sum_{\{S_\alpha\}}^{\pm 1} e^{\frac{1}{2}\beta 
\sum_{\alpha\ne\beta}Q_{\alpha,\beta}S_\alpha S_\beta},
\label{zrep}
\end{eqnarray}
where ${\rm s.p.}$ denotes saddle point over the elements $Q_{\alpha,\beta}$. 
In our case the indexes $\alpha$, $\beta$ take the form 
$\alpha=(s,a)$ $\beta=(u,b)$ with $s,u=1,...,L$, 
$a=1,...,n(s)$ and $b=1,...,n(u)$. The new interesting term 
with respect to the previous analysis is $\zeta[Q]$, that we write as 
\begin{eqnarray}
 \zeta[Q]=\sum_{\{S_a(s) \}}^{\pm 1} e^{\frac{1}{2}\beta 
\sum_{s,u}\sum_{a,b}Q_{a,b}(s,u)
S_a(s) S_b(u)}.
\end{eqnarray}
For small Q one can study the development in powers of Q and many
results can be straightforward obtained by the algebra we have just
derived. In the following we will derive compact expressions that goes
beyond such an expansion.
We then substitute (\ref{RS},\ref{risp}), we  get for $s>u$
\begin{eqnarray}
\sum_{a,b}Q_{a,b}(s,u)
S_a(s) S_b(u)& =& (\sum_a S_a(s))C(s,u)(\sum_b S_b(u))\nonumber\\
&+&(\sum_a S_a(s))R(s,u) S_1(u)du,
\end{eqnarray}
while for $s=u$
\begin{eqnarray}
\sum_{a\ne b}Q_{a,b}(s,s)
S_a(s) S_b(s)=(\sum_a S_a(s))C(s,s)(\sum_b S_b(s)).
\end{eqnarray}
Introducing a field $i{\hat h}(s)ds =\sum_{a=1}^{n(s)}\sigma_a(s)$ and its conjugate 
$\beta h(s)$, and mixing freely discrete and continuous time notation, 
we rewrite 
\begin{eqnarray}
 \zeta[Q]&=&\sum_{\{S_a(s) \}}^{\pm 1}\int \prod_{u}d{\hat h}(u)dh(u)\;
 \exp\left({-\frac{1}{2}\int ds\; du\; \beta {\hat h}(s)C(s,u) {\hat h}(u)
-i\beta \int du\; {\hat h}(u) h(u)}\right)\nonumber\\
&\times& \exp{ \left(\int ds\;du\; i{\hat h}(u)R(u,s)S_1(s)+\sum_u \beta h(u)(S_1(u) 
+\sum_{a=2}^{n(u)}S_a(u))\right)
}\;.
\end{eqnarray}
Notice that the spins $S_a(u)$ in replicas $a>1$ are decoupled, and can be summed over; resulting in terms of the kind $(2 \cosh h(u))^{n(u)-1}\to (2 \cosh h(u))^{-1}$ for $n(u)\to 0$. The final expression is
\begin{eqnarray}
 \zeta[Q]&=&\sum_{\{S_1(s) \}}^{\pm 1} \int \prod_{u} \left( \frac{ e^{\beta h(u) S_1(u)}}{2\cosh(\beta h(u))}d{\hat h}(u)dh(u)\right)
\nonumber\\
&\times& \exp\left({-\frac{1}{2}\int ds\;du\;\beta {\hat h}(s)C(s,u) {\hat h}(u)
-i\beta \int du\;  {\hat h}(u) h(u) +\int ds\;du\; i{\hat h}(u)R(u,s)S_1(s)}\right) .
\end{eqnarray}
Taking into account the first term in (\ref{zrep}), one finds the self-consistency 
equations 
$C(s,u)=\langle S_1(s)S_1(u)\rangle$ and 
$T(u)R(s,u)=\frac{\delta\langle S_1(s)\rangle}{\delta h(u)}$. 

Our findings have a clear interpretation: in the long chain limit 
they describe the
equilibrium of spins with their local field as well as the adiabatic
evolution of the field. In fact, we find that local spin and the field are related by  
\begin{eqnarray}
P(S(t)|h(t))= \frac{ e^{h(t) S(t)}}{2\cosh(h(t))}.  
\label{cond}
\end{eqnarray}
while the field is determined by 
\begin{eqnarray}
h(t)=\eta(t)+\int_0^t ds\; R(t,s) S(s), 
\label{campo}
\end{eqnarray}
where $\eta(t)$ is a zero mean Gaussian variable with covariance 
$\langle \eta(t)\eta(s)\rangle=C(t,s)$. 
The equations are closed observing that in the long 
chain limit, where the dependence of $h(t)$ and $R(t,s)$ on time is slow,
one can substitute $S(s)$ in (\ref{campo}) with its conditional average from (\ref{cond}) $m(s)=\tanh h(s)$.   
Our equations provide the long time limit of
the Eisfeller-Opper equations for the dynamics of the SK model
\cite{opper}. The same equations are the skeleton of the dynamic
cavity equations in the long time limit (as discussed in reference
\cite{MPV}, that can be derived by a direct analysis in
\cite{SOMZIPP}).

It is well known that in the SK model the dynamical formalism allows
to recover many equilibrium quantities related to replica symmetry
breaking, like e.g. the function $q(x)$ that describes the statistics
of pure states \cite{PARISI}.  In dynamics this quantity is intimately
related to the breakdown of fluctuation dissipation theorem \cite{CuKu} 
and the emergence of effective temperatures \cite{CuKuPe}. Our analysis 
unambiguously shows that their appearence is associated to quasi-equilibrium
sampling of phase space.

\section{Perspectives}

In this paper we have formalized the notion that slow glassy dynamics
consists in quasi-equilibrium exploration of configuration space. Our
analysis has been achieved with mean-field models as reference. In the
last thirty years such models have been a precious guide in forming
physical pictures of the glassy behavior of more realistic systems, we
can then conjecture that the quasi-equilibrium description holds in
general for marginally stable glassy dynamics. Indeed we think that
quasi-equilibrium exploration of phase space is at the heart of the
emergence modified fluctuation dissipation relations \cite{CuKu} and
effective temperatures \cite{CuKuPe}, the equivalence between
equilibrium and dynamics discussed in \cite{FMPP} and the time
reparametrization invariance properties \cite{russi}. All these 
properties have been well verified to hold beyond the mean-field level. 

The formalism we have introduced here is very general and open 
new perspectives in the comprehension of glassy dynamics. 

We would like to mention here a few problems where we expect it 
will lead to relevant progress. First of all let us quote some
problems where the chain constraints do not make work, in connection with 
glassy slow time dynamics. 

At the most basic level our formalism suggests how to obtain a
sensible discretization of dynamic equations in the slow time
limit. This could be useful in the context of cavity dynamical method
for spin models on sparse graphs and Bethe lattices, where a direct
dynamical approach leads to hard technical difficulties
\cite{Coolen_etc}. This opens new perspectives for studying dynamical
processes on network and network dynamics \cite{saad}, such as
epidemic or damage spreading.

At a more fundamental level, one can tackle the task of computing
dynamical quantities that could not be computed in a dynamical
approach. For example, the dynamical entropy, which is related to the
terms of order $n(L)$ of the free-energy.

The basic property that slow dynamics is a quasi-equilibrium state,
suggests to use our pseudo-dynamic formalism to study dynamical 
quantities in cases where a direct dynamic approach
is problematic. For example we argue that it enables in principle to
study slow dynamics (alpha dynamical processes) of liquids, starting
from equilibrium approaches based on the replica method. A remarkable 
example is the HNC approximation, where one can find equations 
similar to Mode Coupling Theory, but in
the context of a fully consistent theory. Work is on the way 
in this direction \cite{FPU}. 

On the side of systems with different temperatures, one can use the
formalism to follow metastable states adiabatically  in temperature 
\cite{FloLenka}. A related question concerns the inclusion of replica
symmetry breaking effects.  Finally we would like to mention interpretation
questions that are far from be settled, to start with the surprising
appearance of the dynamical response in a purely thermodynamic
setting.

The other class of problems which can be addressed 
is the one where the constraints are effective. 
The chain construction constitutes a powerful probe of configuration space.  
With working constraints one can study free-energy barriers among states. 
The properties of these barriers and whether the barriers 
of pseudo-dynamics are related to the 
ones of the real dynamics is a question to be investigated.  

{\bf Acknowledgments} We thank F. Krzakala and L. Zdeborova for pointing
our attention to the chain construction, J. Kurchan, F. Ricci-Tersenghi and 
P.-F. Urbani for discussions.
 
The research of S.F. has received funding from the European Union,
Seventh Framework Programme FP7-ICT-2009-C under grant agreement n. 265496.

\end{document}